\documentclass[journal]{vgtc}                
\ifpdf
  \pdfoutput=1\relax                   
  \usepackage{graphicx}                
  \DeclareGraphicsExtensions{.pdf,.png,.jpg,.jpeg} 
\else
  \usepackage{graphicx}                
  \DeclareGraphicsExtensions{.eps}     
\fi%

\graphicspath{{figures/}{pictures/}{images/}{./}} 

\usepackage{microtype}                 
\PassOptionsToPackage{warn}{textcomp}  
\usepackage{textcomp}                  
\usepackage{mathptmx}                  
\usepackage{times}                     
\usepackage{cite}                      
\usepackage{tabu}                      
\usepackage{booktabs}                  

\usepackage{soul}
\makeatletter
 \def\SOUL@hlpreamble{%
 \setul{}{2ex}
 \let\SOUL@stcolor\SOUL@hlcolor
 \SOUL@stpreamble
 }
\makeatother

\onlineid{0}

\vgtccategory{Research}
\vgtcpapertype{please specify}

\title{A Design Space for Surfacing Content Recommendations in Visual Analytic Platforms}

\author{Zhilan Zhou, Wenyuan Wang, Mengtian Guo, Yue Wang, and David Gotz}
\authorfooter{
\item
 Zhilan Zhou is with the Department of Computer Science at the University of North Carolina at Chapel Hill. E-mail: zzl@email.unc.edu.
\item
 Wenyuan Wang is with the School of Information and Library Science at the University of North Carolina at Chapel Hill. E-mail: vaapad@live.unc.edu.
\item
 Mengtian Guo is with the School of Information and Library Science at the University of North Carolina at Chapel Hill. E-mail: mtguo@email.unc.edu.
\item
 Yue Wang is with the School of Information and Library Science at the University of North Carolina at Chapel Hill. E-mail: wangyue@unc.edu.
\item
 David Gotz is with the School of Information and Library Science and the Department of Computer Science at the University of North Carolina at Chapel Hill. E-mail: gotz@unc.edu.
}

\abstract{Recommendation algorithms have been leveraged in various ways within visualization systems to assist users as they perform of a range of information tasks. One common focus for these techniques has been the recommendation of content, rather than visual form, as a means to assist users in the identification of information that is relevant to their task context. A wide variety of techniques have been proposed to address this general problem, with a range of design choices in how these solutions surface relevant information to users. This paper reviews the state-of-the-art in how visualization systems surface recommended content to users during users' visual analysis; introduces a four-dimensional design space for visual content recommendation based on a characterization of prior work; and discusses key observations regarding common patterns and future research opportunities.
} 


\CCScatlist{ 
 \CCScat{K.6.1}{Management of Computing and Information Systems}%
{Project and People Management}{Life Cycle};
 \CCScat{K.7.m}{The Computing Profession}{Miscellaneous}{Ethics}
}

\vgtcinsertpkg

\begin{document}



\colorlet{transcolor}{yellow!50}
\colorlet{transcolor}{yellow!0}
\sethlcolor{transcolor}

\firstsection{Introduction}
\maketitle

Visual analytics as a discipline was born from the recognition that the process of deriving meaningful insights from large volume, high dimensional, heterogeneous, and often conflicting sources of data is fraught with an array of challenges \cite{thomas_illuminating_2005}. Much of the research in this area is motivated by complex information problems in which users must work to discover relevant data elements, to view those data from different perspectives, and to progressively connect disparate pieces of information to build an understanding of their analytical target as part of an ongoing sensemaking process \cite{pirolli2005sensemaking}.

Reflecting on the scale of the challenge in terms of both volume and complexity of information, visual analytics techniques generally aim to combine computational techniques with human analysis activities in a cooperative fashion to exploit the complementary strengths of both computer and human
\cite{sacha2014knowledge}.  Moreover, the proliferation of machine learning has led to a wide variety of proposals for how to leverage mixed-initiative and data-driven algorithms during visual analysis \cite{endert2017state}. 

Perhaps not surprisingly, researchers have experimented with many ways in which the combination of algorithmic and human approaches could best be leveraged. This includes, for example, ``user-in-the-loop'' approaches in which users can provide various forms of guidance to the underlying algorithms (e.g., \cite{bernard2017comparing,stolper2014progressive}).  It also includes ``analytics-in-the-loop'' approaches in which model-driven algorithms are designed to provide various forms of assistance to human analysts. 

This latter category--using computational approaches to help a human user make analytical progress--can itself come in many forms. One frequent approach is to algorithmically recommend specific \emph{visualizations} (e.g., \cite{moritz2019formalizing,hu2019vizml}). In this formulation, the algorithmic problem is to determine the ``best'' view of a given set of data in a particular task context.  Alternatively, and more relevant to the work presented in this paper, computational approaches have also been proposed to recommend specific \emph{content} to users.  In this problem, the goal is to algorithmically help users more effectively identify new information that is relevant to their analytic context.

Visual analytics systems that support this type of content recommendation must include several different components. First, the system must maintain some representation of a user's current analytic context.  Second, the system must include some form of recommender which, given the representation of a user's context, evaluates new information to identify which content should be recommended.  Finally, recommended content must be surfaced to the user through the user interface of the visual analytics system.  

It is this last element---the way in which recommended content is surfaced to users during visual analysis---that is the focus of this paper. By design, content recommendation algorithms are used to identify new content relevant to the user's analytic context which they might not identify on their own. This makes content recommendation a potentially powerful tool in support of the sensemaking process.  
However, the way in which recommended content is incorporated into a user's analytic workflow can influence the utility of the recommended content itself. The utility is heavily influenced by the design choices of how, when, and where that content is presented to the user.

To better understand the design space for the content recommendation within visual analytics platforms, and motivated by our own work on this topic \cite{zhou_modeling_2021}, we conducted a formal survey of the visual analytics literature to identify prior efforts at content recommendation. We searched the literature for adaptive visualization systems that algorithmically identify and surface new content with the aim of helping users discover additional relevant information given their analytical context. 

Based on an analysis of the identified articles, a four-dimensional design space is proposed that captures the primary design choices for communicating recommended content during visual analysis. The papers found in the search are then characterized using the proposed design space to identify common patterns, rare design choices, and potential opportunities for future research.

In this way, the research contributions of this paper include: (1)~a systematic review of the literature to identify the state-of-the-art in how visualization systems surface recommended content during visual analysis; (2)~a design space for visual content recommendation interfaces; and (3)~identification of both common approaches and opportunities for future research.

The remainder of this paper is organized as follows. A brief review of related work is provided in Section \ref{sec:related} to better frame the scope of our literature survey.  Section \ref{sec:methodology} then introduces the methodology behind the survey including the search strategy.  The design space is presented in Section \ref{sec:design_space} and used to characterize the papers identified in the literature search. Finally, before concluding, a discussion in Section \ref{sec:discussion} identifies key patterns, limitations, and opportunities for future research. 

\section{Related Work}
\label{sec:related}

The literature survey and design space presented in this paper focus on the ways in which content recommendations are surfaced to users during visual analysis.  This topic is closely related to two broader areas of research that have received significant attention in recent years.  These include both (a) visualization recommendation systems, and (b) more wide-ranging research on recommender systems within the human-computer interaction (HCI) community.

\subsection{Visualization Recommendation}

Algorithmic solutions that recommend or automatically construct visual representations of data have been studied for several decades (e.g., \cite{roth_interactive_1994}). Broadly speaking, these types of visualization recommendation systems will automatically configure and suggest one or more visual representations for a given set of data based on either data properties, a user's task context (including, potentially, user preferences), or both. Given these inputs, a visualization recommendation system will typically aim to determine the best visualizations to use with the given dataset in order to help the user achieve some analytical goal \cite{vartak_towards_2017}.

A variety of techniques have been proposed to accomplish this type of visualization recommendation, ranging from rule or pattern-based approaches that infer visualization requirements from user activity (e.g., \cite{gotz_behavior_2009}) to machine learning models trained on large numbers of example visualizations (e.g., \cite{hu2019vizml,moritz2019formalizing}).

Despite the variety in both the input and the computational approaches through which visualization recommendations are generated, these systems share a common goal: given an input dataset and perhaps some additional information about the user's task or preferences, recommend one or more views of the input data. These approaches suggest \emph{forms} or \emph{styles} of visualization, but do \emph{not} aim to inject any new \emph{content} into a user's analysis~\cite{bao2022recommendations}. 
In sharp contrast, the focus of this paper is on the content recommendation during visualization rather than visualization recommendation as we have just defined it.  That is, we focus on papers that (a) describe recommendations of new content that should be brought to the user's attention, and (b) how those recommendations are surfaced to the user.  For this reason, visualization recommendations papers (such as those cited in this subsection) are considered outside of the scope for this paper.

\subsection{Recommendation Systems in HCI}
\label{sec:related_rec_in_hci}

The HCI research community has an extensive record of impactful research exploring different types of recommendation systems in application scenarios that extend well beyond visualization. Fundamentally, the term \emph{recommendation system} (or \emph{recommender system}) is used to describe any system that makes content suggestions to a user based upon a computed measure of relevance between said content and user’s interest~\cite{pazzani2007content}. 

Usually, recommender technology is provided in combination with other information access techniques, such as query-based search~\cite{cremonesi_user_2017}. For example, recommendation systems are used widely on websites such as search engines or shopping websites to recommend related documents to read (e.g.,~\cite{beel2016paper}) or related products to purchase (e.g.,~\cite{smith2017two}), respectively. They are also widely used on online social media platforms to recommend videos, posts, and other content based on both user's current interactive activity and models of users' historical preference (e.g., \cite{covington2016deep}).

Many different approaches to solving the content recommendation problem have been proposed including content-based filtering methods, collaborative filtering methods, and hybrid methods. Research on these topics has often focused on optimizing quantitative measures of recommendation accuracy \cite{shani2011evaluating}.  However, researchers have also recognized the critical importance of user experience and interaction design \cite{knijnenburg2012explaining}. This has, for example, let to suggested ``design patterns'' for user interfaces within the context of recommender systems \cite{cremonesi_user_2017}.  The contributions of this paper are similarly focused on content recommendation and the ways in which these recommendations are surfaced to users.  However, unlike past work, this paper aims to specifically study ways in which content recommendations are surfaced within visual analytics platforms.

\section{Methodology}
\label{sec:methodology}

The design space proposed in this paper is derived from an analysis of articles identified through a formal literature survey. This section provides an overview of the survey strategy and the analysis process used to inform the design space. The numbers of articles found at each stage of the search process are also reported.

\subsection{Search and Screening}
\label{sec:search}

The literature search performed for this paper followed a search strategy and screening process that adheres to the recommended procedures specified in the PRISMA standard (Preferred Reporting Items for Systematic Reviews and Meta-Analyses)~\cite{liberati2009prisma}. PRISMA defines a multi-stage procedure that has been widely adopted in health and medical fields as a ``best practice'' for systematic literature reviews. PRISMA also defines a standardized flow diagram to depict the various stages and the number of articles considered, included, and/or excluded at each step.

The PRISMA flow diagram for the literature survey in this paper is shown in Figure~\ref{fig:prisma}. As shown in the diagram, articles were identified from two different sources: IEEE Xplore \cite{ieee_xplore} and the ACM Digital Library \cite{acm_dl}. A full-text keyword search was performed within both online digital libraries to identify candidate articles. Given the focus of the search, ``adaptive visualization(s)'' was the phrase used as search terms. 
The keywords were free-text and not selected from a controlled vocabulary. Searches were conducted in both IEEE Xplore digital library and ACM digital library in December 2021.

This choice of keywords, \hl{selected to reflect the focus of this survey on methods of visually surfacing recommended content within visualization-based interfaces, }was made after experimentation with various alternatives in both the IEEE and ACM digital libraries, as well as informal exploratory searches made through Google Scholar \cite{google_scholar}.  Alternatives explored included, for example, ``visualization recommendation.'' However, this phrasing is widely used to describe systems that recommend the form of visualization rather than the content (see Section~\ref{sec:related}.1).  As a result, it returned articles that were out of scope and missed many important papers that deserved to be included.  Alternatively, a search for ``content recommendation'' was excessively broad due to the generic terms used.  This search therefore returned an enormous number of articles, many of which were not related to content recommendation at all as defined in this paper (much fewer content recommendations in the context of visualization). 

The choice of scoping the search to IEEE and ACM digital libraries reflects the central role these two repositories have in archiving peer-reviewed high-quality articles relevant to visualization and recommendation research.  Together, these two resources index a broad range of technical journals and conference proceedings, and they include a number of the leading publications in the disciplines relevant to the scope of this search.

In total, 175 articles were identified via the specified keywords. A screening process was followed to remove duplicates and exclude irrelevant articles by examining titles and abstracts.  This process resulted in a total of 92 articles for potential inclusion in the survey.

\hl{The choices of keywords and digital libraries used in the search clearly have a significant impact on the set of articles. The impacts of these choices are discussed in more detail along with other limitations of our approach in Section~{\ref{sec:limitations}}}.

\subsection{Eligibility Assessment}

The 92 articles that emerged from the screening process were inspected more closely for eligibility via a full-text assessment. This step ensured that papers which appeared relevant from the title and abstract screening were indeed appropriate for inclusion based on the research topic.  This further narrowed the number of candidate articles to 46. 

Once articles were determined to be relevant in terms of topic, two additional eligibility constraints were applied which excluded 23 additional articles. \hl{First, papers were required to present a working prototype to ensure that our survey focused on designs that were used in actual implementations from prior work. This requirement eliminated 12 articles that proposed theoretical concepts or computational algorithms\mbox{\cite{bai_context_2013,banos_mining_2015,burkhardt_innovations_2020,deuschel_influence_2018,domik_user_nodate,golemati_context-based_2006,guo_real-time_2017,ruotsalo_interactive_2014,schonhage_flexible_1997,jianwei_yu_framework_2009,pombinho_chameleon_2011,nazemi_measuring_2014}},
and 8 papers that included results from user studies without prototypes of interactive visualizations\mbox{\cite{barral_understanding_2020,barral_effect_2021,conati_exploring_2008,kim_bayesian-assisted_2021,oscar_towards_2017,steichen_user-adaptive_2013,toker_individual_2013,wu_learning_2021}}. 
Second, because content recommendation is a dynamic process that occurs while a user interacts with a system, papers that met the prototype requirement were also required to describe user interaction. This requirement resulted in the elimination of 3 additional papers\mbox{\cite{chaudhuri_self-adaptive_2009,miyamura_adaptive_2011,yu_parallel_2007}}.}

\hl{Finally, we removed 6 papers\mbox{\cite{rusu_adaptive_nodate,zhang_adaptive_2018,coors_resource-adaptive_2002,wasinger_scrutable_nodate,ahn_envisioning_2008,wiza_periscope_2004}} where the prototypes were deemed not relevant to the scope of the study, were described with insufficient detail to include, or  were identified as duplicates that described the same prototypes as other papers in our collection. These steps reduced the included set to 17 papers.}

\subsection{Additional Search Following Citation Links}
\label{sec:citations}

Recognizing the limitations of keyword search, the research team enriched the collection of relevant documents by examining all citations from the 17 papers which emerged from the eligibility determination process. Every document cited in the references section for these 17 documents went through the same screening and eligibility steps as the original articles found in the IEEE and ACM digital libraries.  This process resulted in 9 additional articles being included for a total of 26 articles in the final collection.

\subsection{Reference Management}

\begin{figure*}[thb]
\centering  
\includegraphics[width=0.6\textwidth]{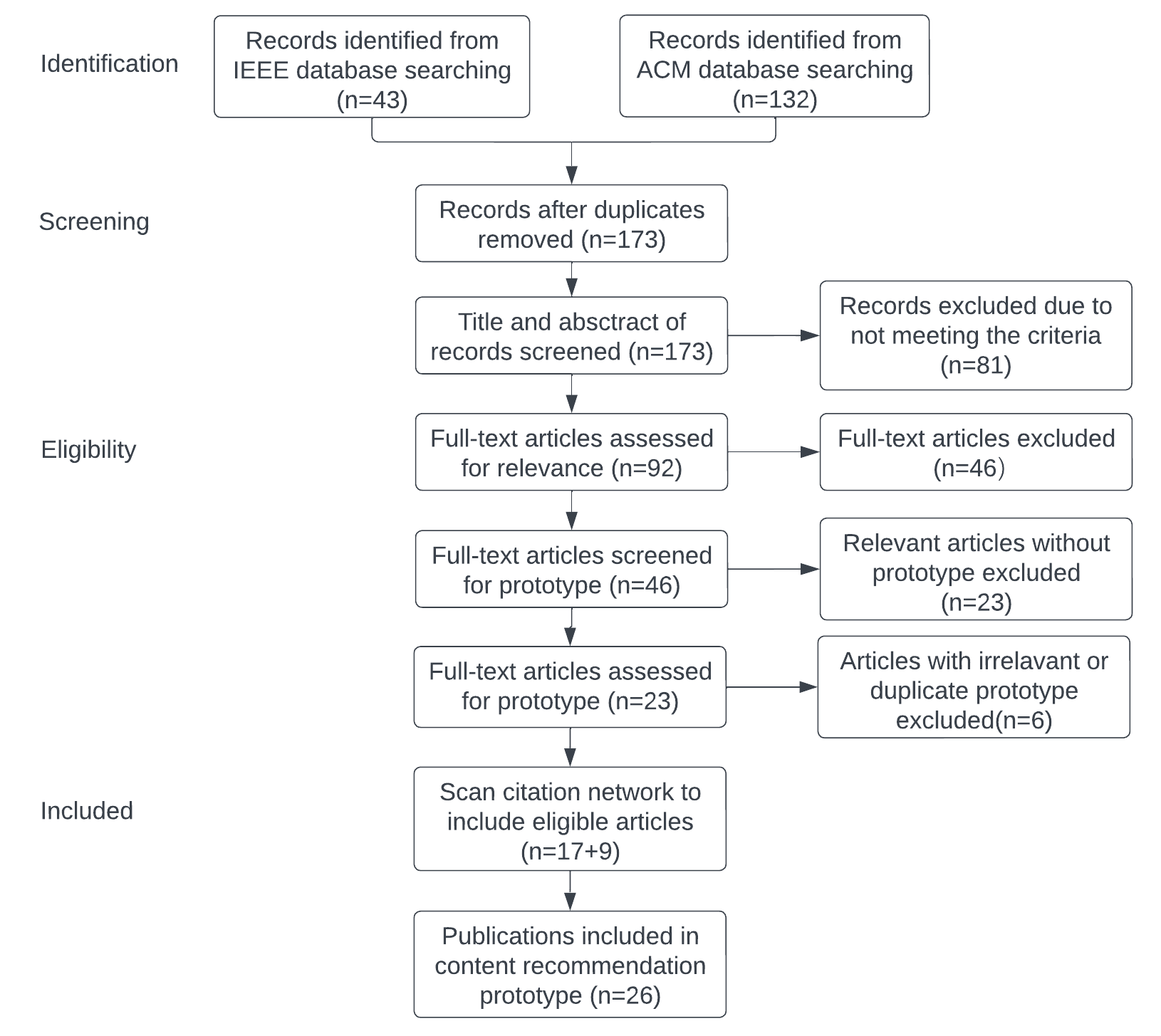}
\caption{The PRISMA (Preferred Reporting Items for Systematic Reviews and Meta-Analyses) flow diagram for our searching and screening process.}
\label{fig:prisma}
\end{figure*}

The open-source reference management software Zotero was leveraged to organize all the references identified in the search and screening process. In addition, spreadsheets were maintained to keep track of the overall review process, decisions, and statistics, as reported in Figure~\ref{fig:prisma}.

\subsection{Meta-Analysis Process}
\label{sec:process}

The final set of 26 publications that were included in the survey was then analyzed by the authors of this paper to extract details regarding the ways in which the described prototypes surfaced recommended content to users. 
Descriptors of various techniques used in prototype systems in the literature were recorded in a spreadsheet \hl{following an open coding process. The descriptors were iteratively refined and grouped to identify the key aspects that could be used to characterize various design choices. New values were added as needed to ensure that all identified prototypes could be accurately described. As the descriptors were refined, previously characterized papers were revisited to ensure they were labeled using the latest set of descriptors. }
At the end of this process, the descriptors eventually consolidated into the four-dimensional design space described in Section~\ref{sec:design_space}.

As the final design space emerged, two authors (ZZ and WW) revisited all articles included in the search and formally characterized each article using terms from the design space. The two authors worked independently and then resolved disagreements through discussion and resolution. In rare circumstances, disagreements led to minor revisions of how dimensions in the design space were described.  When this occurred, previously characterized papers were revisited to ensure they were consistently analyzed.  The final results of this process are illustrated in Figure~\ref{figure:papers} and discussed in detail throughout the remainder of this paper.
\section{The Design Space}
\label{sec:design_space}

This section introduces a four-dimensional design space which provides a framework for describing how content recommendation capabilities are made visible to users within visualization systems.  As described in Section~\ref{sec:methodology}, the design space was derived from an analysis of relevant papers identified through a systematic review of the research literature.  This section first introduces the dimensions of this design space, and then uses the design space to categorize and discuss prior research identified in the literature review.  


\subsection{Dimensions of the Design Space}
\label{sec:def}

Informed by our analysis of the publications identified in our literature search, we define four key dimensions that help characterize the interface design for content recommendation within visual analytic systems: \textbf{Directness}, \textbf{Forcefulness}, \textbf{Stability}, and \textbf{Granularity}.
This subsection formally defines each of these dimensions. 

\hl{We note that we explicitly consider as out of scope issues related to: (a)~the identification of what content should be recommended (e.g., similarity measures, individual vs. collaborative  approaches, etc.); and (b)~how to model user interests and behaviors as the basis for computing recommendations. These out-of-scope aspects are essential parts of recommender systems even when visualization is not employed~\mbox{\cite{jannach2010recommender}}, and they are key aspects to much of the related work referenced in Section~{\ref{sec:related_rec_in_hci}}. However, they are beyond the scope of the design space proposed in this paper because they relate to what content is recommended. In contrast, the proposed design space is focused on the methods via which the recommended content is made visible to users.}

\begin{figure}[!ht]
\centering    
	\includegraphics[width=3in]{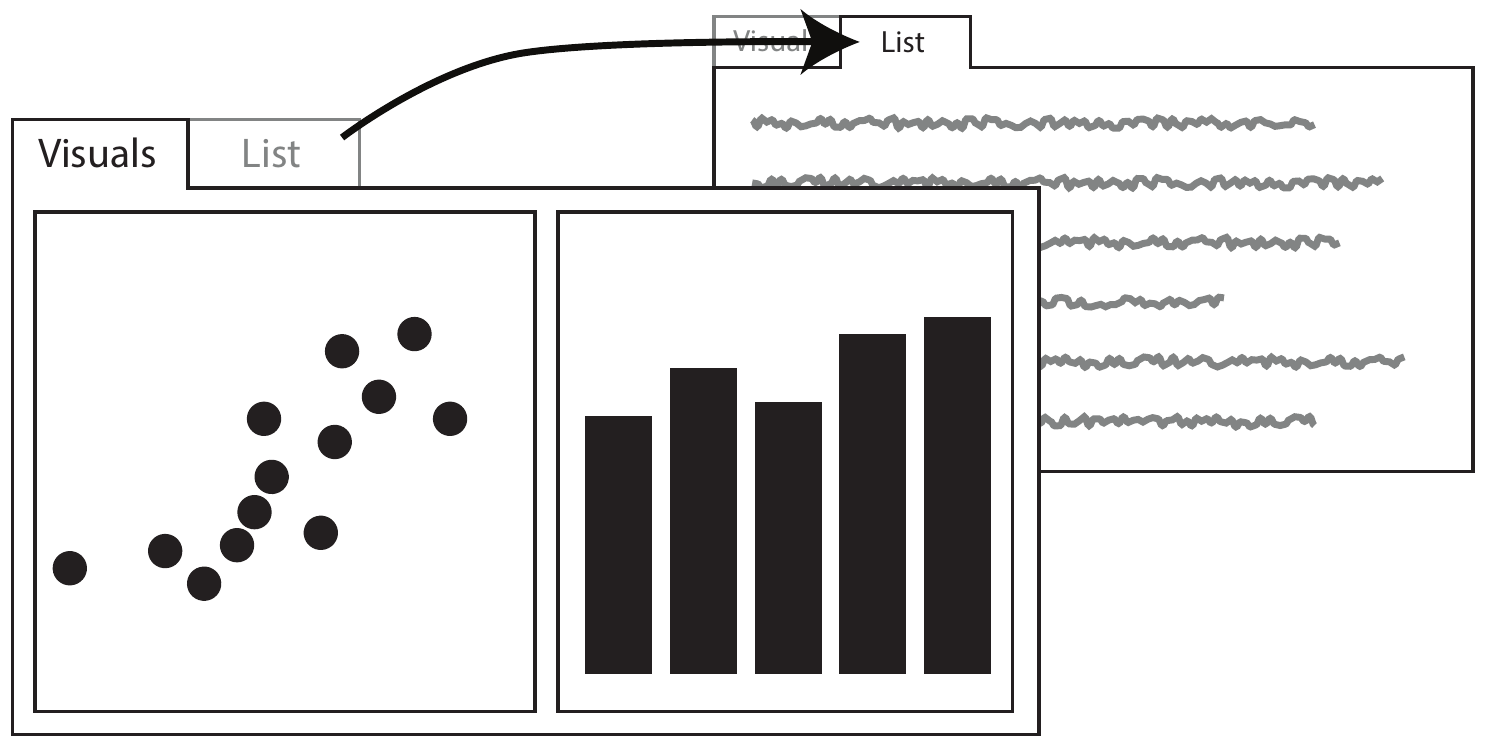}
	\caption{A sketch of HVAP, the Hypothetical Visual Analytics Platform used as an example when defining the design space proposed in this paper. The hypothetical interface includes two tabs: one with visualizations and another with a text-based list.} 
	\label{figure:hvap}
\end{figure}

To help readers understand how these dimensions correspond to practical design decisions, we provide concrete examples using a \underline{h}ypothetical \underline{v}isual \underline{a}nalytic \underline{p}latform, \textit{HVAP}. This hypothetical system has a relatively simple interface with just two tabs: the first tab contains visualizations in the form of a scatterplot and a histogram; the second tab contains a text-based list. This is illustrated in Figure~\ref{figure:hvap}. HVAP is used as an example for explanatory purposes throughout the remainder of this section. 

\subsubsection{Directness}

The \emph{directness} dimension captures the relationship between where a user is interacting with a visualization and the location at which recommended content is surfaced. The directness dimension takes a binary value, which can be either \textit{Direct} or \textit{Indirect}.

A \emph{direct} design places recommended content within the same interface component with which the user is interacting. For example, a direct recommendation for a user interacting with the scatterplot in HVAP would appear within the same scatterplot.

In contrast, content surfaced via an \emph{indirect} recommendation would be visible through another interface component.  For example, for a user interacting with the HVAP scatterplot, indirect recommendations might be surfaced via the histogram (in the same tab but different component) or via the text list (in a different tab).

\subsubsection{Forcefulness}

The \emph{forcefulness} dimension represents how intrusively the recommendation is surfaced with respect to the user's workflow. Unlike directness, forcefulness exists on a scale with different design choices exhibiting levels of forcefulness.  Based on the analysis of the articles identified in our literature, forcefulness is assessed on a five-point scale with 1 being the least forceful and 5 being the most forceful.  

In discussing forcefulness, we use the term \textit{context} to refer to the information being displayed within a visual analytics system before any recommended content is surfaced.  
When the context is displaced completely by a recommendation, the user is compelled to attend to the recommendation before returning to their prior analytical workflow. For example, if a user is interacting with the scatterplot in HVAP and a recommendation is displayed such that it obscures the visualizations entirely, this would forcefully interrupt a user's analysis. We rank this type of design as Level~5 on the forcefulness scale. 

In contrast, a fully passive design for the recommendation will not alter the context in any way and will only surface information in out-of-view locations for users to optionally attend  to at some time in the future. For example, recommended content for the same HVAP user interacting with the scatterplot could be displayed within the text list.  We rank such a design as Level~1.

Table~\ref{tab:forcefulness} summarizes the full forcefulness scale using the Level~5 and Level~1 scenarios as the extremes of the scale. As the table shows, two factors contribute to the forcefulness of a recommendation: the degree to which context is displaced, and the visibility of the recommended content.

\begin{table}[t]
\begin{center}
\begin{tabular}{@{}lll@{}}
\toprule
Forcefulness Level             & Visible Context & Visible Recommendation \\ \midrule
5 (Most Forceful) & None   & Recommendation only    \\
4                 & Part            & Highlighted over context \\
3                 & All             & In context, highlighted \\
2                 & All             & Parallel to the context \\
1 (Most Passive)  & All             & None (Hidden)          \\ \bottomrule
\end{tabular}
\vspace{0.15cm}
\caption{The forcefulness scale has five levels and depends on two factors: the degree to which the context is displaced, and the visibility of the recommended content.}
\label{tab:forcefulness}
\end{center}
\end{table}

Example recommendation scenarios within HVAP for each level of forcefulness are provided in Table~\ref{tab:forcefulness-eg}. The examples all assume users are interacting with the HVAP scatterplot at the time of recommendation.

We emphasize that while directness and forcefulness can seem similar, they are  conceptually different dimensions. Using HVAP as an example, if a user selects a point in the scatterplot, recommended content can be shown through a tooltip that appears near the hovered point (direct, forcefulness = 4), or a highlight of the five closest nodes and a dimming of others (direct, forcefulness = 3). Alternatively, recommended content can be surfaced in the bar chart component, which is outside of the user's current focus in the scattterplot (making this alternative indirect). The interface can highlight relevant bars with respect to the selected point (indirect, forcefulness = 3), or replace the original bar chart with a new one (indirect, forcefulness = 5). 



\begin{table*}[th]
\begin{center}
\begin{tabular}{@{}lll@{}}
\toprule
Forcefulness Level & Elements Visible to User                               & Recommendation Location                      \\ \midrule
5 (Most Forceful)  & Only the recommendation           & The visualization tab, obscuring all context \\
4                  & Just the scatterplot and the recommendation                    & Visualization tab, obscuring some context \\
3                  & Scatterplot, histogram, and the recommendation                       & Visualization tab, highlighted within the scatterplot \\
2                  & Scatterplot, histogram, and the recommendation                       & Visualization tab, shown in parallel to context \\
1 (Most Passive)   & Scatterplot and histogram; recommendation not visible & List tab (not visible without additional user interaction)                                \\ \bottomrule
\end{tabular}
\vspace{0.2cm}
\caption{Examples of different levels of content recommendation forcefulness for a user interacting with the scatterplot in HVAP (see Figure~\ref{figure:hvap}). The level of forcefulness depends on both (a) what information is visible to users, and (b) the location of the recommendation.}
\label{tab:forcefulness-eg}
\vspace{0.1cm}
\end{center}
\end{table*}


\subsubsection{Stability}

The \emph{stability} dimension describes the temporal dynamics of when the content being recommended is updated within the interface.  In this dimension, we only consider updates to the visible portion of a recommendation system. The underlying algorithm that prepares the content may update its output on its own schedule that is independent of when the content in the interface is updated, and this algorithmic timing is not the focus of this design space. Design choices within the stability dimension can be classified into three general categories: periodic, event-driven, and on-demand policy. 

\emph{Periodic} updates occur at a fixed time interval. For example, the display of recommended content could be revised every 3 seconds.  \emph{Event-driven} updates occur in response to some underlying system event. For example, the display of recommended content could be revised every time a relevant document appears in a data stream, or in response to a resizing of a user interface window. Event-driven designs often result in interfaces that exhibit irregularly-timed changes to the display of recommended content. Both periodic and event-driven designs result in interface updates that are system initiated.  These designs results in less-stable interfaces which can change the visual display at times that are not expected and/or inconvenient to a user.

In contrast, \emph{on-demand} systems provide the most stable experience because they only update the display of recommended content in response to explicit user requests (e.g., clicking on a button to update recommendations). This design can result in fewer updates and longer delays between the algorithmic identification of relevant content, but offers more agency to users and reduces the chances that updates will disrupt a user's workflow.

\subsubsection{Granularity}

The \emph{granularity} dimension describes the unit of content that is recommended to the user. \hl{By ``unit of content'' we refer to the atomic unit of information that is being recommended or visualized (e.g. a document; a semantic concept; a dimension of a dataset; a recommended database entity such as the data for a single house in a real estate application). To characterize this aspect, we focus on the difference between (a) the unit of content that a user is viewing within the context of the recommendation, and (b) the unit of content being recommended. The granularity of recommended content can be coarser, similar, or finer compared to the context.} 

In practice, recommendations often contain heterogeneous content with mixed granularities. For example, imagine that HVAP is used to analyze food pricing strategies in various countries. In this scenario, imagine the scatterplot is showing the relationship between the wholesale and market prices of various food categories (e.g., fruit, dairy, grains) for a specific country. 
If the HVAP system recommends information about the prices of oranges, we would consider this a finer granularity recommendation because an orange is a narrower concept than the fruit category. If the system recommends information about the prices of citrus fruits in another country, we would consider this a similar granularity recommendation. If the system instead recommends market research data for both individual fruits as well as the overall fruit category, it would be a mixed granularity recommendation (finer and similar).


\begin{figure*}[!ht]
\centering    
	\vspace{0.3cm}
	\includegraphics[width=0.9\linewidth,trim=10 10 10 10]{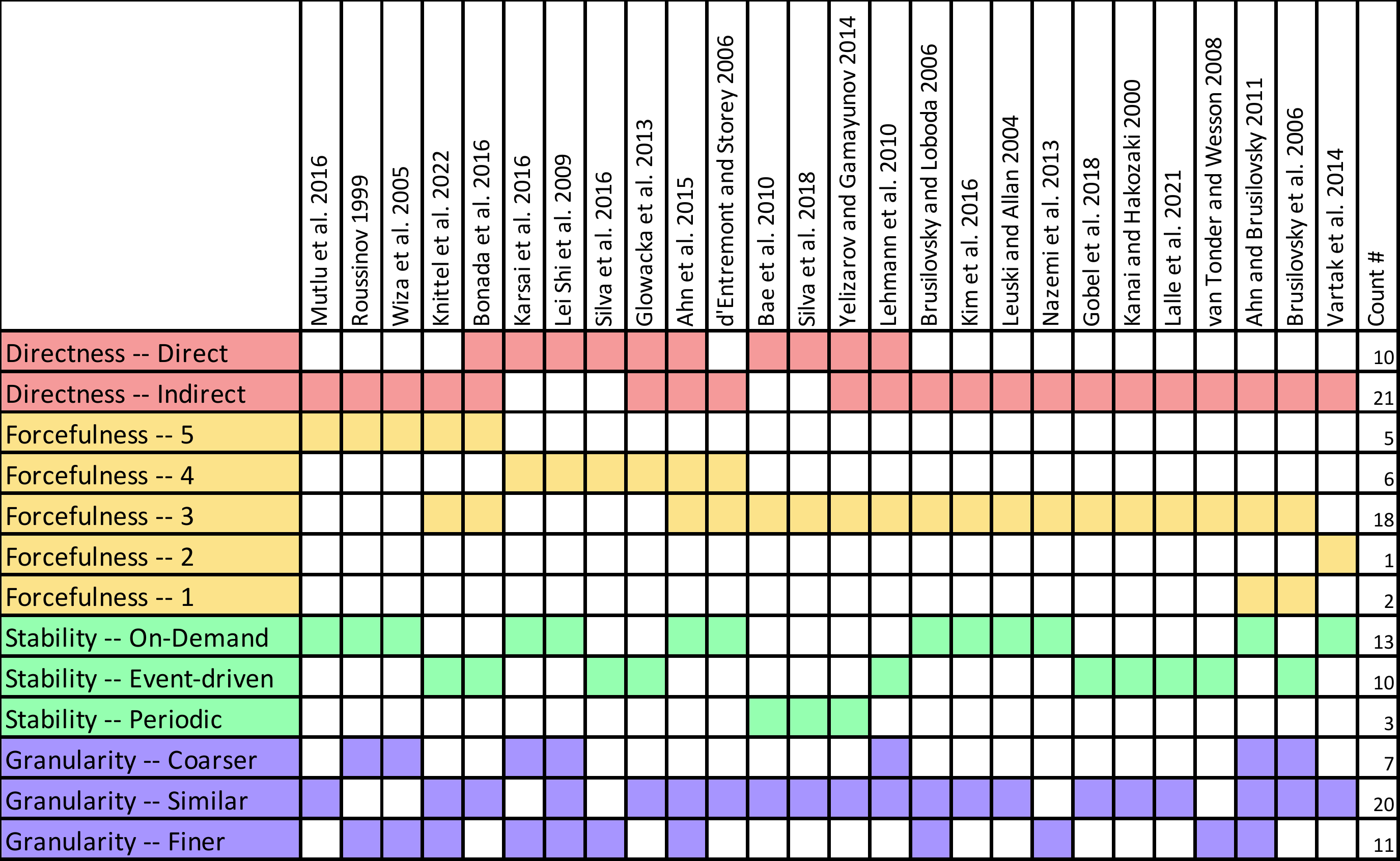}
	\vspace{0.1cm}
	\caption{A systematic review of the literature found a total of 26 papers as illustrated in Figure~\ref{fig:prisma}. Each paper was reviewed by two authors and classified within the proposed four-dimensional design space for the content recommendation within visual analytics platforms. This table summarizes the findings for all papers across the four design space dimensions: directness, forcefulness, stability, and granularity.} 
	\label{figure:papers}
	\vspace{0.2cm}
\end{figure*}

\subsection{Mapping Literature in the Design Space}
\label{sec:applying}
 
Given the design dimensions defined above, we used the design space to characterize each of the 26 papers identified in the literature review. The findings are reported by dimension, with the full results summarized in the table in Figure~\ref{figure:papers}. The results show significant variation cross all dimensions, but also some interesting patterns that speak to design trade-offs for visualization interfaces that adopt content recommendation capabilities.

\subsubsection{Directness}
\label{sec:directness}

The \textbf{directness} dimension describes the location at which a recommendation is displayed. Designs that surface recommendations within the interface component that is being interacted with by a user adopt a \textit{direct} design, while those that show recommendations in other interface components are considered \textit{indirect}. Often, systems provide recommendations in more than one way. For example, in HVAP, a user interacting with the scatterplot might see recommendations both within the scatterplot (direct) and text list (indirect).  This can result in designs that exhibit a combination of both direct and indirect recommendations within different parts on the platform.

{\bf Direct Designs.} A total of ten papers \cite{bonada_personalized_2016,ahn_personalized_2015,glowacka_directing_2013,yelizarov_adaptive_2014,lehmann_interactive_2010,karsai_clustering_2016,lei_shi_himap_2009,silva_visual_2016,bae_supporting_2010,silva_leveraging_2018} in our survey incorporate at least one form of direct recommendation. These designs highlight the recommendation content within the interface component that is the focus of user interaction.

A concrete example of this approach can be found in the work of Bae et al. \cite{bae_supporting_2010}. They developed an extensible multi-application platform that included a document visualization as the primary interface to support document triage. While users examine the visualization to discover details about various documents, the platform will highlight potentially relevant documents within the visualization to direct users' attention to partial or whole documents that best match users' interests as inferred by the system. Because the highlighted recommendations appear within the same visualization that the users are exploring, this is an example of direct recommendation design.

{\bf Indirect Designs.} A total of 21 papers \cite{bonada_personalized_2016,ahn_personalized_2015,glowacka_directing_2013,yelizarov_adaptive_2014,lehmann_interactive_2010,knittel_real-time_2022,mutlu_vizrec_2016,roussinov_internet_1999,wiza_method_2005,dentremont_using_2006,ahn_guiding_2011,brusilovsky_wadein_2006,brusilovsky_adaptive_2006,gobel_improving_2018,kanai_browsing_2000,kim_generating_2016,lalle_gaze-driven_2021,leuski_interactive_2004,nazemi_adaptive_2013,van_tonder_using_2008,vartak_seedb_2014} identified in our review of the literature demonstrated at least one form of indirect recommendation. These systems included design elements which surfaced recommended content through interface components that were outside the user's current focus of interaction.

For example, Kim et al. \cite{kim_generating_2016} describe a system that recommends re-expressions of spatial measurements (e.g. 78 miles) from texts in a way that contextualizes the distances in terms that are more familiar to a user (e.g. 8 times the distance between your home and your company). A user's primary focus of interaction in this system is the text itself, but the personalized spatial analogies are shown visually using interactive maps placed beside the text.  This placement of the recommendation in a visualization beside and distinct from the main area of interaction makes this an example of indirect recommendation.

Most often, indirect recommendations are located within an interface component that is visible and alongside the user's primary area of interaction.  In some rare cases, however, indirect recommendations have been placed on a separate page that requires additional interactions to be revealed. For example, Brusilovsky et al. \cite{brusilovsky_adaptive_2006} presented a system providing personalized recommendations for a structured repository of educational examples. Users would typically interact with the examples and browse the repository list on the primary component of the interface. The system then included a secondary display area which visualized
a 2D layout of recommended concepts (selected from the pre-defined repository).  Users need to perform an additional click to view these recommendations.

Similarly, the educational system proposed by Ahn \& Brusilovsky~\cite{ahn_guiding_2011} uses a secondary interface to share recommendations, which requires an additional click to access. The system described in that paper adopts a primary visualization interface that uses a Self Organizing Map (SOM) to organize educational resources by topic. Users can click to access a list of recommended topics that are displayed on a secondary interface.

{\bf Discussion.} Overall, about two-fifths of the papers in our survey adopt a direct design. In contrast, more than four-fifths of the papers adopt an indirect design. This makes indirect designs approximately twice as common.  Moreover, half of the direct-design systems also provide indirect recommendations.

Direct designs make recommendations more visible by placing them squarely within a user's area of attention. This approach, therefore, may give recommendations more visibility even for focused users immersed in complex cognitive tasks. However, this same attribute of direct designs has the potential to be more disruptive to a user's analytical work.  In addition, because direct approaches must be designed to visually integrate within an existing view, there are more constraints on how the recommendations are visually represented. 

Indirect designs, because they appear outside of the user's primary area of focus, may at times be less visible than their direct counterparts.  However, such designs usually have ample space and freedom of design for how recommendations are displayed. Moreover, a more forceful design can be used to overcome limitations in visibility. We hypothesize that these reasons explain the greater popularity of indirect designs.

\subsubsection{Forcefulness}
\label{sec:forcefulness}

The \textbf{forcefulness} dimension indicates how intrusively the recommendation is displayed, with values ranging from 5 (most forceful) to 1 (least forceful) as defined in Section~\ref{sec:def}.  These ratings are based on two factors which influence the forcefulness of a recommendation: (a) the degree to which a recommendation displaces a user's context, and (b) the visibility of the recommended content.  The five levels, derived from an analysis of the papers collected in our literature review, are summarized in Table~\ref{tab:forcefulness}. 

We note that there are some cases (e.g. \cite{dentremont_using_2006}) where systems recommend content in more than one way, at times with different design choices regarding what contextual information remains visible and how the recommended content is displayed. In these cases, each approach to recommendation is judged independently resulting in some papers being assigned to multiple levels of forcefulness.
The remainder of this subsection describes which papers fit into each of the five levels and provides concrete examples for each.

{\bf Level 5.} This level represents the most forceful methods of presenting recommendations, reflecting designs which require a user to pay full attention to the recommended content.  A total of five papers from our review~\cite{bonada_personalized_2016,knittel_real-time_2022,mutlu_vizrec_2016,roussinov_internet_1999,wiza_method_2005} fit within the Level~5 category for forcefulness. 

Three of these systems are designed to visualize arbitrarily large datasets \cite{roussinov_internet_1999,wiza_method_2005,mutlu_vizrec_2016} through a query-based interaction model. The systems analyze user queries to determine a recommended subset using the most appropriate dimensions of the dataset. Each time the user makes a new request, the visualization is fully refreshed with an entirely new view of new recommended subset of data. 

In other work, a variation of this approach was adopted which used multiple visualization sub-panels. In this design, some sub-panels were used to show
detailed content recommendation while other sub-panels were responsible for visualizing the context \cite{knittel_real-time_2022}. This design allowed recommendations with Level~5 forcefulness which replaced the overall view with new content in some sub-panels, as well as recommendations at lower levels that preserved some context in other sub-panels. 



{\bf Level 4.} This level represents the second-most forceful design choices in our design space as defined in Section~\ref{sec:def}. At this level, unlike Level~5, a portion of the visible context is maintained as the recommendation is surfaced, and the recommended content is displayed in addition to that remaining context.  

A total of six papers~\cite{ahn_personalized_2015,dentremont_using_2006,glowacka_directing_2013,karsai_clustering_2016,lei_shi_himap_2009,silva_visual_2016} from our search reported designs with Level~4 forcefulness.  A common theme is that these systems managed the integration of new recommended content by selectively removing prior content that was deemed least valuable to maintain within the display, while at the same time maintaining other prior content that remained important according to some criteria.

This type of design was commonly observed in visualizations that depict hierarchical structures in some way to manage complexity. For example, Karsai et al.~\cite{karsai_clustering_2016} present a prototype for visualizing provenance data that is represented using graph-based data structures. The system provides tools for users to interactively view, cluster, and simplify provenance structures. The visualized provenance graphs are updated in response to user actions, and each time new recommended nodes can be added while some existing but less relevant nodes will be removed from view. 

Similarly, one of the two designs from d'Entremont and Storey's work \cite{dentremont_using_2006} hides irrelevant nodes and replaced them with a statistically-derived representation of those hidden nodes. This is done to restrict the navigation space for a better focus on recommended nodes. 

This type of design has also been leveraged in systems that concurrently visualize some representation of user models alongside a primary work area. For example, Ahn et al. \cite{ahn_personalized_2015} propose a design that augments the display of recommended keywords with a representation of a model of user interests that is constantly updated in response to user interactions with the overall system. As new recommended keywords are identified and incorporated into the view, less relevant keywords are de-prioritized but remain on display. Previous keywords that have become irrelevant are removed. Similarly, Glowacka et al. \cite{glowacka_directing_2013} propose a prototype
that will hide less relevant keywords from the view with recommended contents.

{\bf Level 3.} This level represents systems which maintain the full current context of a user's visualization while integrating recommendations within that context.  This was the most frequent design approach observed in our literature review with 18 papers describing recommendations that fit within the Level~3 category~\cite{bonada_personalized_2016,knittel_real-time_2022,ahn_personalized_2015,dentremont_using_2006,lehmann_interactive_2010,yelizarov_adaptive_2014,bae_supporting_2010,silva_leveraging_2018,ahn_guiding_2011,brusilovsky_wadein_2006,brusilovsky_adaptive_2006,gobel_improving_2018,kanai_browsing_2000,kim_generating_2016,lalle_gaze-driven_2021,leuski_interactive_2004,nazemi_adaptive_2013,van_tonder_using_2008}.


The most frequently observed design within this category, found in five of the included papers~\cite{bae_supporting_2010,lehmann_interactive_2010,silva_leveraging_2018,yelizarov_adaptive_2014,lalle_gaze-driven_2021}, was the basic highlighting of recommended content within the existing visual display.  As an example, 
the system proposed by Bae et al.~\cite{bae_supporting_2010} for document triage highlights recommended documents within its visualization of multiple documents, while a details panel showing information about a single document remains unchanged. Similarly, Lehmann et al. \cite{lehmann_interactive_2010} highlights recommended concept clusters as users interact with a concept that appear in a visualization of collections of documents.

The second category of recommendation designs in Level~3 takes a similar but opposite approach: reducing the visual salience of less relevant information rather than highlighting relevant content.  A total of four papers  \cite{dentremont_using_2006,gobel_improving_2018,lalle_gaze-driven_2021,yelizarov_adaptive_2014} leverage this approach. For example, one aspect of a system for visualizing tree-based data proposed by d'Entremont and Storey \cite{dentremont_using_2006} reduces the opacity level of less relevant tree nodes. Similarly, Gobel et al.~\cite{gobel_improving_2018} describe a gaze-based system that reduces the opacity of map legends that are less relevant to a user's interests.

We note that these two approaches (highlighting new recommendations and decreasing the salience of less relevant information) are compatible strategies that can be used simultaneously. For example, this is observed in work from Lalle et al.~\cite{lalle_gaze-driven_2021}, as well as Yelizarov and Gamayunov~\cite{yelizarov_adaptive_2014}.

A third common approach, especially with  visualizations that adopt a visual form showing networks or clusters where the position is flexible, uses the positioning of elements to emphasize a recommendation. For example, work from Kanai and Hakozaki \cite{kanai_browsing_2000} as well as Leuski and Allan\cite{leuski_interactive_2004} use positional shifts to move recommended items ``close'' to the user in 3D visualizations. More recently, Nazemi et al.~\cite{nazemi_adaptive_2013} followed a similar philosophy to visualize document relationships. 

Two projects from Ahn and collaborators \cite{ahn_guiding_2011,ahn_personalized_2015} adopt a more interactive approach in which a spatial positioning of recommended content is arranged nearby user-specified ``points of interest,'' or POIs, based on relevance. The POIs can then be moved interactively to trigger updates from the system.

Meanwhile, Brusilovsky et al.'s~\cite{brusilovsky_adaptive_2006} ADVISE II system adopts yet another use of position to highlight recommended content. It uses position to place visual representations of newly recommended information close to previously consumed content that it is related to.  This is intended to make it easy for users to find the specific recommended content they would find most useful based on what past content they wish to continue examining.

{\bf Level 2.} This level represents systems which show recommended content in parallel to the main context. Only one paper in the review adopted this more subtle form of recommendation, making it the least common forcefulness level in our survey. This form of recommendation was observed in work by Vartak et al. \cite{vartak_seedb_2014} on SeeDB, a system that allows users to build visual queries to a database. The system provided recommendations to users by adding them to a stack of possible combinations of visualizations.

{\bf Level 1.} The least forceful level in our design space shows recommendations only through interface components that are not visible to users without additional interaction.  This is the most subtle form of visual content recommendation as it requires users to specifically act before recommended content can be consumed. Two papers from our literature review fall into this category~\cite{ahn_guiding_2011,brusilovsky_adaptive_2006}. Both systems show recommendations on a separate panel of the interface which is normally not visible. Users must click to access the recommended content shown on those panels.

{\bf Discussion.} Our analysis shows that the extremes of the forcefulness scale are relatively rare in practice, with the majority of papers falling within Level~3. Moreover, six of the 14 papers that included more (Levels~4 or~5) or less (Levels~1 and~2) forceful recommendation designs also incorporated Level~3 design elements.

Between the two extremes, most papers adopted more forceful designs.  This includes five papers that used the most extreme Level~5 compared to just three that used any designs that fit within Level~1 or~2. 
This appears to reflect a goal for developers or designers to ensure that users attend to the recommended content. However, especially given the difficulty that systems often have in modeling user information needs to generate  precise recommendations, there appears to be an opportunity to more deeply explore the use of less forceful recommendation designs.

\subsubsection{Stability}

The \textbf{stability} dimension characterizes how often the recommendation contents get updated, impacting the perceived stability of the visual interface for the user. Design choices within this dimension can be classified into three general categories as defined in Section~\ref{sec:def}: periodic, event-driven, and on-demand policy. It is possible, in theory, that systems combine multiple types of update policies to govern when visible recommendations should be refreshed. However, our review of the literature found that, in practice, each paper could be classified into just one category.

{\bf Periodic.} Systems with periodic stability are those that update the display of recommended content following a regular time interval (e.g., every $n$ seconds).  In our review, we found three papers \cite{yelizarov_adaptive_2014,silva_leveraging_2018,bae_supporting_2010} that followed this approach.

Prototypes in this category typically select a time interval for recommended content updates that is fixed to an arbitrary number of seconds. The time interval tends to be small (e.g., less than 10 seconds), and is sometimes determined in part using data collected from users during usability pilot studies. For example, Silva et al.~\cite{silva_leveraging_2018} updates its recommendations every 2 seconds, while Yelizarov \& Gamayunov~\cite{yelizarov_adaptive_2014} updates its recommendations every 5 seconds. 
    
{\bf Event-driven.} This category represents systems that update their recommendations in response to specific events. The triggering events can be either underlying system events or some form of user action (excluding explicit user requests for recommended content). In total, ten of the 26 papers~\cite{bonada_personalized_2016,lalle_gaze-driven_2021,silva_visual_2016,gobel_improving_2018,kanai_browsing_2000,lehmann_interactive_2010,glowacka_directing_2013,knittel_real-time_2022,van_tonder_using_2008,brusilovsky_adaptive_2006} surveyed included event-driven updates to recommended content.

The largest group of paper in this category respond indirectly to user interactions  \cite{lehmann_interactive_2010,glowacka_directing_2013,knittel_real-time_2022,van_tonder_using_2008,brusilovsky_adaptive_2006,kanai_browsing_2000}. For example, this category includes systems that update recommended content in response to user actions such as a change to visualization settings, selecting or inspecting visualized objects, or moving and resizing items. We describe this type of content recommendation as an ``indirect'' response to user activity because the user actions are not specifically designed to request updates to recommended content. Even if content recommendation were removed from a system, the interactions in this category would still make sense as part of the system's interaction design.  This is in contrast with on-demand designs in the stability dimension which we describe later in this subsection. 

Another group of papers in the event-driven category describe prototypes that respond to eye gazes as the triggering event type~\cite{bonada_personalized_2016,lalle_gaze-driven_2021,silva_visual_2016,gobel_improving_2018}.  One example in this category is recent research by Lalle et al.~\cite{lalle_gaze-driven_2021} in a system designed to support text analysis. The system updates recommendations in response to user eye fixations (in this case, a fixation is defined as maintained gaze on the same location for at least 100 milliseconds).  The number of fixations that meet this duration requirement are tracked, and recommendation updates are triggered when a sufficient number of fixations have been detected.

The final type of triggering event we observed is data entry \cite{kanai_browsing_2000}.  Kanai and Hakozaki describe an approach to the content recommendation which maintains a model of user preferences to guide the identification of related content.  The preference model as well as the update of recommended content within the interface are both triggered by a user's data entry actions as they interact with the system.

{\bf On-demand.} The single largest category in the stability dimension is on-demand. This descriptor applies to systems that only update the recommendations made via the user interface in response to explicit user requests for an update (e.g., by clicking an update button).  This pattern was observed in 13 of the 26 papers in our survey~\cite{kim_generating_2016,leuski_interactive_2004,nazemi_adaptive_2013,ahn_guiding_2011,karsai_clustering_2016,lei_shi_himap_2009,roussinov_internet_1999,ahn_personalized_2015,mutlu_vizrec_2016,brusilovsky_wadein_2006,vartak_seedb_2014,dentremont_using_2006,wiza_method_2005}.

{\bf Discussion.} Exactly half of the papers in our literature review adopted an on-demand policy to update recommendations. The on-demand approach is most straightforward and predictable for users, allowing them to control exactly when updates to visualization or other interface elements will occur. This approach, therefore, provides users with the most stable user experience.  However, on-demand approaches can also results in delays in the communication of recommended content.  Event-driven and periodic updates may result in faster updates, but reduce user agency.

It is also important to note the distinction between computational updates which refresh the content that is identified as best for recommendation, and interface updates which refresh the recommended content that is accessible to users via the user interface. Matching the focus of this paper, the characterizations of prior work reported in this section focuses on interface update policies.  In practice, computation updates and interface updates can be decoupled and occur on different schedules.

\subsubsection{Granularity}

The \textbf{granularity} dimension describes the relationship between the unit of content being recommended and the unit of content being interacted with by a user at the time of recommendation. As defined in Section~\ref{sec:def}, our design space includes three distinct ordinal values for the granularity dimension: coarser, similar, and finer.  Many systems recommended content at multiple levels of granularity, and our review found at least one example of every possible combination of the three levels. 

{\bf Coarser Granularity.} Coarser granularity designs provide recommendations at a higher level of abstraction or via a larger unit of information. A total of 7 papers \cite{roussinov_internet_1999,lehmann_interactive_2010,ahn_guiding_2011,karsai_clustering_2016,lei_shi_himap_2009,wiza_method_2005,brusilovsky_adaptive_2006} were found to recommend content at a coarser granularity. 

For example, Roussinov \cite{roussinov_internet_1999} describes a prototype that uses an Adaptive Self-Organizing-Map (ASOM) to model user queries as a set of concepts. While users view collections of documents containing individual concepts, the system can recommend groups of concepts and concept aggregations based on those groups.

In another example, Ahn \& Brusilovsky~\cite{ahn_guiding_2011} proposed a visualization component called \textit{KnowledgeSea} which supports topic-based navigation. It recommends higher-level concepts based on the topic contents and user interests. Along similar lines, Brusilovsky et al.~\cite{brusilovsky_adaptive_2006} describe an adaptive navigation panel which records a user's progress through information resources and, in response, reveals relevant topic recommendations.  These efforts represent two common use cases for recommendation at a coarser granularity: topic-level summarization, and new topic recommendation.

{\bf Similar Granularity.} Designs that recommend content that has a similar level of granularity as data the user is already viewing within the visual interface are describes as having a similar granularity.  This was the most common form of recommendation in our review of the literature, with 20 out of 26 papers falling into this category~\cite{lehmann_interactive_2010,glowacka_directing_2013,vartak_seedb_2014,dentremont_using_2006,kim_generating_2016,leuski_interactive_2004,gobel_improving_2018,knittel_real-time_2022,ahn_guiding_2011,silva_leveraging_2018,bae_supporting_2010,brusilovsky_adaptive_2006,ahn_personalized_2015,yelizarov_adaptive_2014,mutlu_vizrec_2016,kanai_browsing_2000,brusilovsky_wadein_2006,bonada_personalized_2016,lalle_gaze-driven_2021,lei_shi_himap_2009}.

Consider Bae et al. \cite{bae_supporting_2010} as one example. The system provides users with visualizations that depict individual documents,  and recommendations are also provided at the document level.
Similarly, Kim et al. \cite{kim_generating_2016} recommends visual representations of analogous distance descriptors (e.g. a map view showing 8 times the distance between your home and your company) for spatial measurements (e.g. 78 miles). The recommended analogies always have a granularity that is similar to the original expressions. 

{\bf Finer Granularity.} Recommendations that suggest content that is of a smaller size fit within the finer granularity category. A total of 11 papers~\cite{lei_shi_himap_2009,wiza_method_2005,roussinov_internet_1999,ahn_personalized_2015,knittel_real-time_2022,karsai_clustering_2016,ahn_guiding_2011,brusilovsky_wadein_2006,silva_visual_2016,van_tonder_using_2008,nazemi_adaptive_2013} were identified as fitting into the finer granularity category.
As one example, Silva et al. \cite{silva_visual_2016}, describe an approach which builds a hierarchical view of a large data set, and then provides users with recommendations for specific variables or dimensions that yield deeper, lower-level views of the data which move down the hierarchy.

{\bf Discussion.}
Overall, more than three-fourths of the papers recommend contents with similar granularity. This is perhaps unsurprising as it is typically the most natural level of recommendation.  More specifically, the data is already represented at the level of granularity being used to display it within the visual interface.  Therefore no extra effort is required to obtain data at that same granularity.  Contrast this with coarser or finer granularity recommendation approaches. For any system to recommend content at a different (coarser or finer) granularity, additional work is required to retrieve or construct data at the alternative level of detail.  

Similarly, it is often easier to visually integrate similar granularity recommendations into a user interface because visual representations for data may already exist.  If recommended content is coarser or finer, visual integration of recommended content to an existing context within the interface can itself become challenging.

The one exception where changes in granularity become more natural is in the presence of hierarchical data.  The hierarchical structure works naturally with changes in granularity as systems can leverage the parent-child relationships to move between levels of granularity.  

However, while similar granularity designs are the dominant approach, it is also valuable to observe that a large minority of systems (10 of 26) provide recommendations at multiple levels of granularity. In fact, at least one example was found for every possible combination of the three levels of granularity. 
\section{Discussion}
\label{sec:discussion}


\hl{The four-dimensional design space proposed in Section~{\ref{sec:design_space}} contributes a typology for describing strategies in which recommended content is surfaced in a visual analytic system. This brings several key benefits. First, it allows various content display strategies developed for different applications to be concisely expressed using the same vocabulary (i.e., values along each dimension in the design space). This enables researchers to describe and compare specific existing strategies with others.
Second, the design space provides a framework for improving an existing strategy. Once an existing strategy is mapped in the design space, design alternatives can be systematically generated by movements within the dimensions of the design space. }

This section discusses the main patterns that emerge from the application of this design space to the papers found in our literature review.  This section also highlights important limitations in our approach, and identifies opportunities for future work.

\subsection{Common Trends Across Dimensions}
\label{sec:trends}

The results of characterizing the papers found in our literature review using our four-dimensional design space are reported in Section~\ref{sec:applying} and summarized in Figure~\ref{figure:papers}.  From these results, we observed a small number of high-level trends that span across dimensions of the design space.

First, we observed a relationship between forcefulness and stability in which designs with higher levels of forcefulness (Level~4 or Level~5) had more stable update policies.  More specifically, of the 11 papers forcefulness at Level~4 or~5, seven fell into the on-demand category for stability. This rate for on-demand approaches (7 of 11) is far higher than the rate for on-demand approaches in less forceful recommendation designs (just 6 of 15). This suggests that designers of more intrusive recommendations, which can cause greater disruption to a user's workflow, felt the need to offer a more predictable experience for users. 

Similarly, we found an apparent link between systems that offer recommendations are varied granularity with more stable on-demand designs.  More specifically, 7 of 10 papers with multiple levels of granularity fell within the on-demand category for stability. We hypothesize that the motivation for this pattern is that the presence of heterogeneous granularities of recommendation results in a more complex, cognitively demanding information environment, and that this in turn suggests the need for a more stable recommendation workflow.

Returning to forcefulness, we observed that the most common forcefulness category, Level~3, was typically associated with similar granularity of recommendations (16 of 18). These two approaches are highly compatible in that Level~3 forcefulness designs integrate recommendations within the existing visual context.  This can most easily be done with recommended content that is of the same type and granularity as the already visualized information.  Recommendations that are of non-similar granularity, meanwhile, occur more frequently with either very forceful or less  forceful designs.  These extremes of the forcefulness spectrum provide designers with more freedom because the display of recommended content becomes disconnected from the existing visual display.

\subsection{Content Recommendation as User Guidance}

\hl{Content recommendation can be viewed as a specific form of user guidance. Ceneda et al.~\mbox{\cite{ceneda2016characterizing}} define guidance in visual analytics as a dynamic process that aims to help users make progress in their analyses. Content recommendation is typically employed to provide users with new information that is related to their analysis, and therefore fits within the general definition of a user guidance technique.

Ceneda et al. defined a general model which describes alternative approaches for user guidance at a high level. This model included several dimensions including the type of knowledge gap that an approach is intended to resolve,  the inputs and outputs required, and the degree of guidance. Given the more focused scope of our proposed design space (see Section~{\ref{sec:def}}), the dimensions we proposed are largely independent of those defined by Ceneda et al.

However, there is some alignment between our forcefulness dimension and Ceneda et al.'s \emph{degree} dimension~\mbox{\cite{ceneda2016characterizing}}. Ceneda et al. define guidance degree using three characteristic scenarios: orienting, directing, and prescribing. Orienting helps users understand options for navigation without suggesting system-generated recommendations as to which are better options. In our model, recommended content displayed to achieve an orienting function would map to Level 3 forcefulness. In contrast, directing methods communicate system-generated preferences. In our model, this would map to Level 3 or 4 forcefulness depending on how the interface was designed. Meanwhile, guidance that Ceneda et al. describe as prescribing would fit within Level 4 forcefulness in our model. Ceneda et al. also describe fully-automated guidance which in our model would be considered Level 5 forcefulness.}

\subsection{Limitations}
\label{sec:limitations}

As with all studies based on a systematic review of the literature, the findings in this paper are inherently  dependent on the choice of search terms.  As outlined in Section~\ref{sec:search}, the focus of this search was difficult to specify in terms of keywords without being overly narrow or overly broad. The final choice of terms was a compromise that yielded relevant articles without also returning result sets in the tens or hundreds of thousands which would make a full-text review logistically impossible. The choice to scope the search to IEEE Xplore the ACM Digital Library also helped focus the scope of the search, but at the expense of eliminating some potentially relevant journals or proceedings. We attempted to mitigate these limitations by adding an extra step to our search process. As described in Section~\ref{sec:citations}, we included in our search all papers cited by the papers identified in our original keyword search. This did identify a small number of additional papers from beyond the IEEE and ACM archives.
\hl{However, we acknowledge that relevant articles may still fall out of our search (e.g., recommendations for next-step actions in analytical workflows~\mbox{\cite{srinivasan2021snowy,wang2022interactive}}). Nevertheless, we believe that the general nature of our proposed design space means that it can also be applied in principle to characterize many approaches described in papers missed by our search.

Another limitation relates to our use of an open coding process described in Section~{\ref{sec:process}}. 
Alternative coding approaches could have been followed and might have produced somewhat different results. Relatedly, we recognize that the manual tagging of papers within our design space includes some subjective judgements that are necessarily made based only on the authors' reading of the published articles. To mitigate the risk of errors, every paper was reviewed by two people who worked independently. However, mischaracterizations are still possible based on the limited information provided in some articles.}

\subsection{Opportunities for Future Work}

There are several possible directions for future research that builds on the results presented in this paper.  For example, the relationships between dimensions highlighted in Section~\ref{sec:trends} are based on observations from the analysis of the literature.  Experimental studies designed to better understand the potential design tradeoffs between those dimensions would be a valuable next step.

In addition, the results summarized in Figure~\ref{figure:papers} highlight some areas of the design space are under-explored. Perhaps most interestingly, there is relatively little work exploring more ``subtle'' aspects of content recommendation.  This includes low forcefulness designs, and designs with periodic updates in the stability dimension.  These represent opportunities for research exploring potential use cases where these types of designs might be beneficial. For example, it is possible that these subtle design choices could potentially allow for the display of recommended content in a way that supports serendipitous discovery with negligible impact on users' workflows.  

Finally, while this paper has focused on the visual display of recommended content, future studies examining design decisions for the underlying models and algorithms that support content recommendation could be valuable. Moreover, results from such a study might help uncover how those models and algorithms can directly influence the eventual visual presentation of recommended content.  For example, approaches that leverage hierarchical models might encourage visual designs that surface recommendations at multiple levels of granularity.

\section{Conclusion}

In this paper, we explored the design space for surfacing recommended content within visual analytics platforms.  Like visualization recommendations, which aim to recommend the best form of a visualization to users to advance their analytic goals, content recommendation techniques aim to help users discover related information that they might otherwise overlook. More specifically, content recommendation in visual analytics requires technologies which capture some form of model of a user's information needs given their analytic task context, use that model to identify relevant content, and then surface that recommended content within the user interface to communicate those recommendations to users.

Content recommendation can take many forms, and a wide variety of techniques have been proposed to address this general problem. Moreover, the interface for exposing recommended content to users requires a large number of design decisions which can directly impact the usability and utility of the underlying algorithmic recommendation capabilities.  

This paper describes our efforts to develop a better understanding of the design space for how recommended content can be surfaced within visual analytic platforms.  First, we conducted a systematic review of the relevant literature.  We then analyzed the publications identified from that search to develop a multi-dimensional design space for visual content recommendation.  We formally defined four distinct dimensions in our design space--directness, forcefulness, stability, and granularity--and we outlined the various design alternative within each of those dimensions. We then characterized each of the papers identified in our literature review using the four dimensional design space to gain insights into the design choices made in prior work. While the detailed analysis shows a number of key patterns and commonly adopted approaches, it also highlights that the ``best'' design solution is often highly context dependent with each design choice offering it's own benefits under certain conditions.

\acknowledgments{
The research reported in this article was supported in part by a grant from the National Science Foundation (\#1704018).
}
\bibliographystyle{abbrv-doi}

\bibliography{refs,wang,gotz}
\end{document}